\documentclass[jacsat,article]{achemso}

\usepackage{epstopdf}
\usepackage[caption=false]{subfig}
\usepackage[version=3]{mhchem}
\usepackage{amsmath}
\usepackage{bm}
\DeclareMathOperator{\sgn}{sgn}
\DeclareMathOperator{\coef}{coef}

\usepackage[numbers,sort&compress]{natbib}
\bibpunct[, ]{[}{]}{,}{n}{,}{,}
\renewcommand\bibfont{\fontsize{10}{12}\selectfont}

\author{Anthony Caicedo}
\author{Juan C. Morales} 
\author{Carlos A. Arango}
\affiliation[Universidad Ices]
{Department of Chemical Sciences, Universidad Icesi, Cali, Colombia}
\email{caarango@icesi.edu.co}

\title[An \textsf{achemso} demo]{Geometric-algebraic approach to acid-base equilibrium}


\begin{document}

\maketitle

\begin{abstract}
Closed-form expressions for the pH of ideal weak acid and weak acid buffer solutions (and their titration with a strong base) were obtained and analyzed with the aid of computer algebra systems. These expressions are used to evaluate, without the use of numerical approximations, the precision and accuracy of different approximations commonly employed to calculate the pH. The closed-form expression for the pH of a buffer solution is used to obtain an analytical expression for the pH stability when a strong base is added. Finally, it is shown that the pH expressions for all the systems under study can be obtained from an unique closed-form expression in terms of an effective weak acid constant and an effective acid concentration. 

\end{abstract}

\section{Introduction}

An acid is a chemical substance that dissolves in water producing hydronium ion, $\ce{H3O+}$, and the acid conjugate base. Weak acids dissolve partially in water reaching chemical equilibrium. In the state of equilibrium, the concentration of the chemical species is constant. The equilibrium concentrations of the chemical species are given by the equations of chemical equilibrium and the constraints of mass balance and electrical neutrality. The aqueous dissociation of a weak acid involves four chemical species and four mathematical relations between these species. An aqueous weak acid buffer solution is formed by a weak acid and one of its soluble salts. The equilibrium of a weak acid buffer involves four chemical substances and four equations. In principle is possible to obtain the equilibrium concentrations of all the chemical species for the weak acid dissociation or the buffer solution by solving these systems of equations. In practice, these systems of equations involve nonlinear terms making difficult to obtain exact mathematical expressions for the equilibrium concentrations. The algebraic manipulation of the systems of equations for the weak acid dissociation or the buffer solution give cubic equations for the concentration of $\ce{H3O+}$ \cite{Denbigh1981,Burgot2012,Skoog2022}. The equilibrium concentration of the hydronium is obtained by finding the roots of these cubic equations. Although there exists a cubic formula that gives the explicit roots of a cubic equation, it is not practical to use it due to its complexity \cite{Ladd1980,McNaught1986}. The raw use of the cubic formula requires more than 30 arithmetic operations which makes its use impossible for manual calculation \cite{Bellova2018,Randall2002}.  

A quadratic equation is obtained for the hydronium ion if the auto-ionization of water is ignored. Under this approximation, the equilibrium concentration of the hydronium is obtained simply by solving a quadratic equation. The use of the quadratic formula require seven arithmetic operations, which allows its use for manual calculation. The approximation of ignoring water auto-ionization displays a relative error greater than 10\% for weak acids with $K_\mathrm{a}<10^{-6}\,\mathrm{M}$, and at low analytical concentrations of the acid or the salt of the acid, $C_\mathrm{a,s}<10^{-6}\,\mathrm{M}$. Although the concentrations used in the general chemistry laboratory are generally above $10^{-2}\,\mathrm{M}$, there may be situations in chemical analysis where very low concentrations are handled \cite{Denbigh1981,Bellova2018,Randall2002,Skoog2022}.

General purpose computer algebra systems (CAS) allow the algebraic manipulation of equations and inequalities \cite{Albrecht2013}. The use of CAS in mathematical software, with user friendly graphical interfaces, has extended the use of CAS to all fields of knowledge. Examples of free and non-free CAS are Maple \cite{maple}, Wolfram Mathematica \cite{Mathematica}, Maxima \cite{maxima}, and SageMath \cite{sagemath}. Today mathematical software with CAS capability is easily found in research and academic laboratories and classrooms. Complex problems that were considered beyond the scope of the researchers or students can be solved today on the computer by using CAS mathematical software. The mathematical solution of the aqueous dissociation of weak acids or buffer solutions require functions available in mathematical software. These functions are mainly: (i) the simplification and algebraic manipulations of large equations, and inequalities, and (ii) the use of complex analysis, and graphical representation of functions, inequalities and implicit regions. In this work, Wolfram Mathematica are employed to obtain closed-form expressions for the equilibrium concentrations of the chemical species of four aqueous systems: the weak acid dissociation and its titration by a strong base, the buffer solution and its titration by a strong base \cite{Denbigh1981,Levie2001, Burgot2012,Skoog2022}. The cubic equations for the hydronium ion are deeply analyzed, their roots are obtained by the use of the ancient Cardano's formula for the associated depressed cubic equations \cite{Dickson1914,Dickson1922}. The direct result of Cardano's formula are the cube roots of two complex numbers. The polar representation of these cube roots allows to obtain only three real roots. The use of Descartes rule of signs \cite{Curtiss1918}, and Vieta's formulas \cite{Dickson1914,Dickson1922} allows to demonstrate that only one root is positive. The titration curves of using a strong base to neutralize solutions of weak acids and buffers are analyzed by using the same analytical methods. The pH stability is measured as the change of the pH as the strong base is added.  Finally, a general equation for the concentration of $\ce{H3O+}$ is presented. This equation gives the pH for any of the four systems studied in this work in terms of an effective acid constant $k$ and an effective acid concentration $c$. 

\section{Theory and Methods}\label{sec:theory}

\subsection{Weak acids}
The aqueous dissociation equilibrium of a weak acid $\ce{HB}$ is given by the chemical equations 
\begin{align}
    \ce{HB + H2O &<--> H3O+ + B-},\label{ce:HA_diss}\\
    \ce{2 H2O &<--> H3O+ + OH-}.\label{ce:H2O_auto}
\end{align}
Relevant chemical species are $\ce{H3O+}$, $\ce{OH-}$, $\ce{HB}$, and $\ce{B-}$ with molar concentrations  $\ce{[H3O+]}$, $\ce{[OH^-]}$, $\ce{[HB]}$, and $\ce{[B^-]}$ \cite{Denbigh1981,Burgot2012,Skoog2022}. 

A low concentration solution of the acid $\ce{HB}$ is prepared in water at analytical concentration $C_\mathrm{a}$. Once the system reaches chemical equilibrium, the concentrations of the chemical species are given by four physical conditions: the weak acid dissociation constant, $K_\mathrm{a}$, the water auto-ionization constant, $K_\mathrm{w}$, the electric neutrality, and the mass balance,
\begin{align}
    K_\mathrm{a}&=\frac{\ce{[H3O+]}\ce{[B^-]}}{\ce{[HB]}},\label{eq:Ka}\\
    K_\mathrm{w}&=\ce{[H3O^+]}\ce{[OH^-]},\label{eq:Kw}\\
    \ce{[H3O+]}&=\ce{[OH^-]}+\ce{[B^-]},\label{eq:charge_balance}\\
    C_\mathrm{a}&=\ce{[B^-]}+\ce{[HB]},\label{eq:matter_balance}
\end{align}
respectively. Acid constants $K_\mathrm{a}$ have units of concentration and their value range typically between $10^{-10}$ and $10^{-1}$ $\mathrm{M}$. It is convenient to define the dimensionless variables and constants ${x=\ce{[H3O^+]}/\sqrt{K_\mathrm{w}}}$, ${y=\ce{[OH^-]}/\sqrt{K_\mathrm{w}}}$, ${w=\ce{[HB]}/\sqrt{K_\mathrm{w}}}$, ${z=\ce{[B^-]}/\sqrt{K_\mathrm{w}}}$, ${c_{\mathrm{a}}=C_\mathrm{a}/\sqrt{K_\mathrm{w}}}$,  ${k_{\mathrm{a}}=K_\mathrm{a}/\sqrt{K_\mathrm{w}}}$, and ${k_\mathrm{w}=1}$. In terms of these dimensionless variables and constants, equations \eqref{eq:Ka}--\eqref{eq:matter_balance} are replaced by
\begin{align}
    k_\mathrm{a}&=\frac{xz}{w},\label{eq:ka}\\
    k_\mathrm{w}&=xy=1,\label{eq:kw}\\
    x&=y+z,\label{eq:cb}\\
    c_\mathrm{a}&=w+z.\label{eq:c}
\end{align}
The combined use of equations \eqref{eq:kw}$-$\eqref{eq:c} gives the equations ${z=x-\frac{1}{x}}$ and ${w=c_\mathrm{a}+\frac{1}{x}-x}$. The use of these equations and the physical constraints on the concentrations $x$, $w$ and $z$, gives, after some algebraic manipulation, the mathematical conditions
\begin{align}
    x&>0,\\
    x^2&>1, \\
    -x^2+c_\mathrm{a} x+1&>0.
\end{align}
Algebraic manipulation of these inequalities leads to the range of the dimensionless concentration $x$
\begin{equation}\label{eq:concentration_x_weak_acid}
    1<x<\tfrac{1}{2}\left(c_\mathrm{a}+\sqrt{c_\mathrm{a}^2+4}\right).
\end{equation}
The left inequality gives the condition of acid pH, $[\ce{H3O+}]>10^{-7}\,\textrm{M}$, the right inequality gives the strong acid dissociation limit.

The combined use of equations \eqref{eq:kw}$-$\eqref{eq:c} in equation \eqref{eq:ka} gives
\begin{equation}\label{eq:ka_equation}
    k_\mathrm{a}=\frac{x\left(x-\tfrac{1}{x}\right)}{c_\mathrm{a}-\left(x-\tfrac{1}{x}\right)}.
\end{equation}
Algebraic manipulation on this equation gives the cubic equation $P_\mathrm{a}=0$, with the cubic polynomial 
\begin{equation}\label{eq:polynomial_P}
    P_\mathrm{a}=x^3+k_\mathrm{a}x^2-\left(1+c_\mathrm{a}k_\mathrm{a}\right)x-k_\mathrm{a}.
\end{equation}
The coefficients of $P_\mathrm{a}$ are given by the 4-tuple
\begin{equation}\label{eq:coefficients_Pa}
\begin{split}
    \coef[P_\mathrm{a}]&=\left(a_3,a_2,a_1,a_0\right)\\
    &=\left(1,k_\mathrm{a},-\left(1+c_\mathrm{a}k_\mathrm{a}\right),-k_\mathrm{a}\right).
\end{split}
\end{equation}
The signs of the coefficients $ \coef[P_\mathrm{a}]$ are given by
\begin{equation}\label{eq:signs_coeff_Pa}
\begin{split}
    \sgn[P_\mathrm{a}]&=\left(\sgn a_3,\sgn a_2,\sgn a_1, \sgn a_0\right)\\
    &=\left(+,+,-,-\right).
\end{split}
\end{equation}
Descartes' rule of signs says that the number of positive roots of a polynomial is, at most, equal to the number of sign changes of its ordered list of coefficients \cite{Curtiss1918}. The use of Descartes's rule on the sequence \eqref{eq:signs_coeff_Pa} indicates that only one root is positive. The full characterization of the roots of $P_\mathrm{a}=0$ is given by the discriminant of the polynomial \eqref{eq:polynomial_P}, $\Delta[P_\mathrm{a}]$ \cite{Irving2004,Dickson1914,Dickson1922}, 
\begin{equation}\label{eq:discriminant_x}
    \Delta[P_\mathrm{a}]=4\left(c_\mathrm{a}^3k_\mathrm{a}^3+\left(k_\mathrm{a}^2-1\right)^2+c_\mathrm{a} k_\mathrm{a}\left(5k_\mathrm{a}^2+3\right)\right)+c_\mathrm{a}^2k_\mathrm{a}^2\left(k_\mathrm{a}^2+12\right).
\end{equation}
Clearly $\Delta[P_\mathrm{a}]>0$, hence the cubic equation $P_\mathrm{a}=0$ must have three distinct real roots $\rho_i$, with $i=1,2,3$ and $\rho_1<\rho_2<\rho_3$. Since $x=\ce{[H3O^+]}/\sqrt{K_\mathrm{w}}$ is related to a molar concentration, only one of the roots $\rho_i$ is expected to be a positive number. Although Descartes' rule has already shown that there must be only one positive root, a further analysis is necessary to find the local extremes of $P_\mathrm{a}(x)$. These extremes are given by the solutions of $P'_\mathrm{a}(x)=0$. The derivative of $P_\mathrm{a}(x)$ is given by
\begin{equation}
    P'_\mathrm{a}(x)=3x^2+2k_\mathrm{a} x-\left(1+c_\mathrm{a}k_\mathrm{a}\right), 
\end{equation} 
which has discriminant $\Delta[P_\mathrm{a}']=4\left(k_\mathrm{a}^ 2+3c_\mathrm{a}k_\mathrm{a}+3\right)$.
The roots of $P_\mathrm{a}'(x)=0$ are
\begin{equation}
    x_{1,2}=\tfrac{1}{3}\left(-k_\mathrm{a}\mp\tfrac{1}{2}\sqrt{\Delta[P_\mathrm{a}']}\right).
\end{equation}
Since $\Delta[P_\mathrm{a}']>0$, it is easily seen that $x_1<0$. Algebraic manipulation on the inequality $x_2>0$ leads to $1+c_\mathrm{a}k_\mathrm{a}>0$, corroborating that $x_2>0$. Since $P_\mathrm{a}(x)$ is a cubic polynomial and the coefficient of the cubic term is positive, it must be given that $\lim_{x\to\pm\infty} P_\mathrm{a}(x)=\pm\infty$, therefore $x_1$ must be a local maximum and $x_2$ a local minimum, with $P_\mathrm{a}(x_1)>0$, and $P_\mathrm{a}(x_2)<0$. The three roots of $P_\mathrm{a}=0$ then obey the inequalities $\rho_1<x_1<\rho_2<x_2<\rho_3$. These inequalities give $\rho_1<0$, and $\rho_3>0$. The use of Vieta's theorem allows to determine the sign of $\rho_2$ \cite{vanderWaerden1991,Virberg2003}, these formulas for $P_\mathrm{a}(x)=0$ are \cite{Dickson1914,Dickson1922}
\begin{align}
    \rho_1+\rho_2+\rho_3&=-k_\mathrm{a},\\
    \rho_1\rho_2+\rho_1\rho_3+\rho_2\rho_3&=-\left(1+c_\mathrm{a}k_\mathrm{a}\right),\\
    \rho_1\rho_2\rho_3&=k_\mathrm{a}.
\end{align}
The use of the last of the formulas gives $\rho_2<0$, therefore only $\rho_3$ is positive. 

An explicit formula for $\rho_3$, in terms of the constants, is obtained by solving the associated depressed cubic equation, \emph{i.e.}, a cubic equation with no quadratic term \cite{vanderWaerden1991,Virberg2003}. The change of variable $x=\bar{x}-\frac{k_\mathrm{a}}{3}$ allows to write the cubic equation $P_\mathrm{a}=0$ as the depressed cubic equation
\begin{equation}\label{eq:depressed_cubic}
      \bar{x}^3+p\bar{x}+q=0,
\end{equation}
with real coefficients
\begin{align}
    p&=-\frac{k_\mathrm{a}^2}{3}-c_\mathrm{a} k_\mathrm{a}-1,\\
    q&=\frac{2k_\mathrm{a}^3}{27}+\frac{c_\mathrm{a}k_\mathrm{a}^2}{3}-\frac{2k_\mathrm{a}}{3},
\end{align}
discriminant $-4p^3-27q^2$ equal to $\Delta[P_\mathrm{a}]$, and  ${p=-\Delta[P_\mathrm{a}']/12}$. 

The coefficient $p$ is evidently negative, meanwhile $q$ can have positive or negative values. Algebraic manipulation on the inequality $q<0$ gives $k_\mathrm{a}^2+\tfrac{9}{2}c_\mathrm{a}k_\mathrm{a}-9<0$, which can be rearranged to obtain $c_\mathrm{a}<\frac{2}{9k_\mathrm{a}}(k_\mathrm{a}^2-9)$. The last inequality implies that negative values of $q$ are obtained with $k_\mathrm{a}<3$, and ${c_\mathrm{a}<\frac{2}{9k_\mathrm{a}}\left(9-k_\mathrm{a}^2\right)}$. Very weak acids, \emph{e.g.} HCN ($k_\mathrm{a}=6.2\times 10^{-3}$) or HOCl ($k_\mathrm{a}=0.4$), would have negative $q$ at $\mu\mathrm{M}$ concentrations. 

Vieta's substitution, $\bar{x}=y-\frac{p}{3y}$, on the depressed cubic equation \eqref{eq:depressed_cubic} gives
\begin{equation}\label{eq:vieta_depressed_cubic}
    y^3-\frac{p^3}{27 y^3}+q=0.
\end{equation}
The denominator of the second term on the right hand side is simplified by multiplying equation \eqref{eq:vieta_depressed_cubic} by $y^3$, obtaining 
\begin{equation}\label{eq:vieta_6}
    y^6+q y^3-\frac{p^3}{27}=0,
\end{equation}
which is equivalent to the quadratic equation $\xi^2+q\xi-\tfrac{p^3}{27}=0$, in the variable $\xi=y^3$, with roots 
\begin{equation}
\begin{split}\label{eq:roots_vieta_6}
        \xi_{1,2}&=-\frac{q}{2}\pm\sqrt{\frac{27q^2+4p^3}{108}}\\
        &=-\frac{q}{2}\pm \frac{i}{2}\sqrt{\frac{\Delta[P_\mathrm{a}]}{27}}.
\end{split}
\end{equation}
Since $\Delta[P_\mathrm{a}]>0$, the roots $\xi_{1,2}$ are complex conjugates, $\xi_1=\xi_2^*$. The definition $\zeta=\xi_1$ allows to use $\zeta$ and $\zeta^*$ as the roots of \eqref{eq:vieta_6}. It is convenient to use the polar representation of $\zeta$, ${\zeta=\|\zeta\|e^{i\theta}}$ with 
\begin{equation}\label{eq:z_modulus}
\begin{split}
    \|\zeta\|&=\frac{1}{2}\sqrt{q^2+\frac{\Delta[P_\mathrm{a}]}{27}}\\
    &=\left(\tfrac{1}{6}\sqrt{\Delta[P_\mathrm{a}']}\right)^3,\\
\end{split}
\end{equation}
and 
\begin{equation}\label{eq:theta_def}
    \theta=\arctan{\left(-\frac{q}{2},\frac{\sqrt{\Delta[P_\mathrm{a}]}}{6\sqrt{3}}\right)},
\end{equation}
with $\theta\in(0,\pi)$ as the angle between $\zeta$ and the positive real axis on the Argand plane. The angle $\theta$ is related to the trigonometric solution obtained by Nickalls for the roots of the cubic equation \cite{Nickalls1993}. 

The roots of the depressed cubic equation are obtained from $\alpha=\sqrt[3]{\zeta}$ and $\beta=\sqrt[3]{\zeta^*}$. The cubic roots $\alpha$ and $\beta$ have three values each, $\alpha_n$ and $\beta_n$,
\begin{align}
    \alpha_n&=\sqrt[3]{\|\zeta\|}\exp{\left(i\left(\frac{\theta}{3}+\frac{2n\pi}{3}\right)\right)},\\
    \beta_n&=\sqrt[3]{\|\zeta\|}\exp{\left(i\left(-\frac{\theta}{3}+\frac{2n\pi}{3}\right)\right)},
\end{align}
with $n=0,1,2$. 


Since the cubic equation \eqref{eq:depressed_cubic} has three real roots, the addition of two cubic roots $\alpha_n+\beta_m$ must give a real number. This is possible only if $\operatorname{Im}(\alpha_n)=-\operatorname{Im}(\beta_m)$.
There are only three possible combinations that fulfill this requirement: 
\begin{align}
    \alpha_0+\beta_0&=2\sqrt[3]{\|\zeta\|}\cos{\left(\frac{\theta}{3}\right)},\\
    \alpha_2+\beta_1&=-2\sqrt[3]{\|\zeta\|}\cos{\left(\frac{\theta+\pi}{3}\right)},\\
    \alpha_1+\beta_2&=-2\sqrt[3]{\|\zeta\|}\sin{\left(\frac{\theta+2\pi}{3}\right)}.
\end{align}
The roots of the depressed cubic equation \eqref{eq:depressed_cubic} are
\begin{align}
    \bar{r}_1&=\alpha_1+\beta_2,\label{eq:rbar_1}\\
    \bar{r}_2&=\alpha_2+\beta_1,\label{eq:rbar_2}\\
    \bar{r}_3&=\alpha_0+\beta_0,\label{eq:rbar_3}
\end{align}
with $\bar{r}_1<\bar{r}_2<\bar{r}_3$. 

The roots of $P_\mathrm{a}=0$ are given by
\begin{align}
    \rho_1=\bar{r}_1-\tfrac{k_\mathrm{a}}{3},\label{eq:rho_1}\\
    \rho_2=\bar{r}_2-\tfrac{k_\mathrm{a}}{3},\label{eq:rho_2}\\
    \rho_3=\bar{r}_3-\tfrac{k_\mathrm{a}}{3}.\label{eq:rho_3}
\end{align}
By using Vieta's formulas it was shown that $\rho_3$ is the only positive root. The positive root $\rho=\rho_3$ is explicitly given by
\begin{equation}\label{eq:rho3_q_negative}
\begin{split}
    \rho&=2\sqrt[3]{\|\zeta\|}\cos{\left(\theta/3\right)}-\tfrac{k_\mathrm{a}}{3}\\
    &=\tfrac{2}{3}\sqrt{k_\mathrm{a}^2+3c_\mathrm{a}k_\mathrm{a}+3}\cos{\left(\theta/3\right)}-\tfrac{k_\mathrm{a}}{3}.
    \end{split}
\end{equation}
The weak acid and low concentration limits give $\lim_{c_\mathrm{a}\to 0}{\rho}=1$, and  $\lim_{k_\mathrm{a}\to 0}{\rho}=1$. The strong acid limit gives 
\begin{equation}
\lim_{k_\mathrm{a}\to\infty}{\rho}=\tfrac{1}{2}\left(c_\mathrm{a}+\sqrt{4+c_\mathrm{a}^2}\right).      
\end{equation}

The concentration $\ce{[H3O+]}$ is given by $\ce{[H3O+]}=\sqrt{K_\mathrm{w}}\rho$, and the pH is
\begin{equation}\label{eq:pH_weak_acid}
\begin{split}
    \mathrm{pH}&=-\log_{10}{\frac{\ce{[H3O+]}}{C^\circ}}\\
    &=7-\log_{10}{\rho},
    \end{split}
\end{equation}
with $C^{\circ}=1\,\mathrm{M}$. The low concentration and weak acid pH limits are given by $\lim_{c_\mathrm{a}\to 0}{\mathrm{pH}}=7$, and  ${\lim_{k_\mathrm{a}\to 0}{\mathrm{pH}}=7}$. The strong acid limit can be used to show that the pH can be negative for $c_\mathrm{a} > 10^{7}$, $C_\mathrm{a}>1\,\mathrm{M}$. Equation \eqref{eq:pH_weak_acid} is useful to calculate the pH of a weak acid at low concentrations (${1-10\,\mu}$M). College books of chemical analysis \cite{Skoog2022} resort to several approximations to calculate the pH of a weak acid in order to avoid the numerical difficulties of solving the cubic equation $P_\mathrm{a}=0$. First, assuming that ${\sqrt{K_\mathrm{w}}\ll\ce{[H3O+]}}$, \emph{i.e.} ${1\ll x}$, in equation \eqref{eq:ka_equation}, it is obtained a quadratic equation for $x$ with solution ${x_\mathrm{approx}^{(1)} =\tfrac{1}{2}\left(-k_\mathrm{a}+\sqrt{k_\mathrm{a}^2+4c_\mathrm{a}k_\mathrm{a}}\right)}$. Further, by using ${1\ll x\ll c_\mathrm{a}}$ in equation \eqref{eq:ka_equation}, it is obtained second (less exact) approximation, ${x_\mathrm{approx}^{(2)}=\sqrt{c_\mathrm{a}k_\mathrm{a}}}$. The relative error of the approximate concentrations $x_\mathrm{approx}^{(i)}$, is given by
\begin{align}
    E^{(i)}=\frac{x_\mathrm{approx}^{(i)}-\rho}{\rho},
\end{align}
with $i=1,2$. It can be shown that the errors $E^{(i)}$ have extreme values
\begin{align}
    \lim_{c_\mathrm{a}\to 0}E^{(i)}&=-1,\\
    \lim_{c_\mathrm{a}\to\infty}E^{(i)}&=0,
\end{align}
for $i=1,2$. 

\begin{figure}[htb]
    \centering
    \includegraphics[scale=0.5]{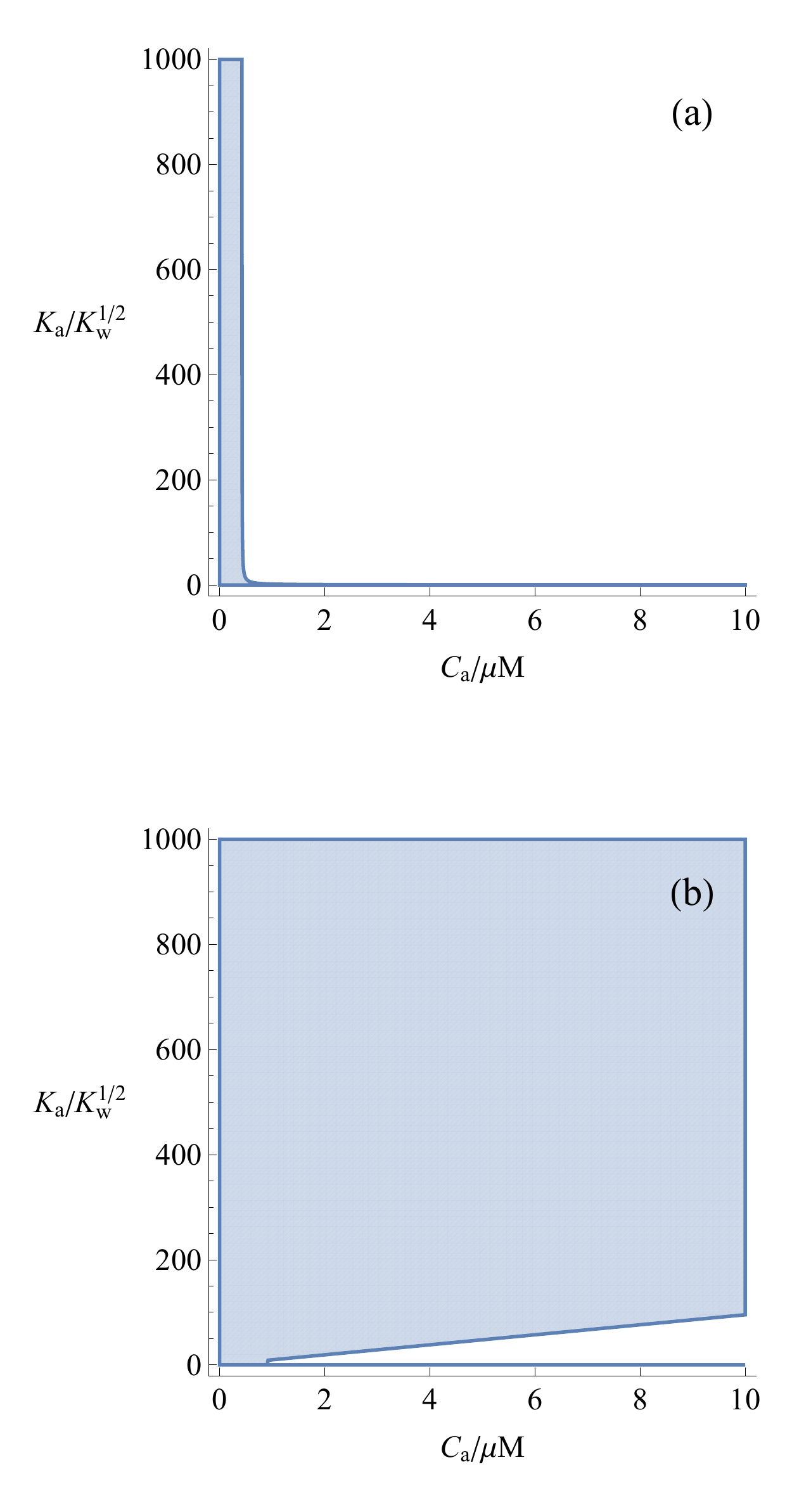}
    \caption{The rule of 5\%. In blue,  regions with relative error (a) $\left|E^{(1)}\right|>0.05$, and (b) $\left|E^{(2)}\right|>0.05$. The white color indicate the region on the $C_\mathrm{a}$-$K_\mathrm{a}$ plane where (a) $x_{\mathrm{approx}}^{(1)}$,  or (b) $x_{\mathrm{approx}}^{(2)}$, are good approximations.}    
    \label{fig:x_approximations}
\end{figure}

Figure \ref{fig:x_approximations}(a) displays the region of the $C_\mathrm{a}$-$K_\mathrm{a}$ plane for which $\left|E^{(1)}\right|>0.05$, figure \ref{fig:x_approximations}(b) shows the region of the $C_\mathrm{a}$-$K_\mathrm{a}$ plane for which $\left|E^{(2)}\right|>0.05$. It is clear that the error of the first approximation, $x_\mathrm{approx}^{(1)}$, is very small even at $\mu\mathrm{M}$ concentrations, this approximation fails for $C_\mathrm{a}<1\,\mu\mathrm{M}$. On the other hand, the approximation $x_\mathrm{approx}^{(2)}$ works well for small acid dissociation constants, $K_\mathrm{a}<10^{-6}\,\mathrm{M}$, and large values of the acid concentration, $C_\mathrm{a}>2\,m\mathrm{M}$. 

The combined use of equations \eqref{eq:kw} and \eqref{eq:cb} give the equilibrium concentration for $z$
\begin{equation}\label{eq:z_of_rho}
    z=\frac{\rho^2-1}{\rho},
\end{equation}
which can be used to obtain the degree of dissociation of the acid $\mathcal{D}_\mathrm{a}=z/c_\mathrm{a}$ \cite{Skoog2022}, which can be rearranged to obtain $c_\mathrm{a}$ as function of $\rho$ and $\mathcal{D}_\mathrm{a}$
\begin{equation}\label{eq:degree_of_dissociation_1}
    c_\mathrm{a}=\frac{\rho^2-1}{\rho\mathcal{D}_\mathrm{a}}.
\end{equation}

The combined use of equations \eqref{eq:ka}--\eqref{eq:c}, the substitution $x=\rho$, and the use of the definition of $\mathcal{D}_\mathrm{a}$, give after some algebra
\begin{equation}\label{eq:ka_in_terms_of_da_and_rho}
    k_\mathrm{a}=\frac{\rho\mathcal{D}_\mathrm{a}}{1-\mathcal{D}_\mathrm{a}}.
\end{equation}

\begin{figure}[htb]
    \centering
    \includegraphics[scale=0.7]{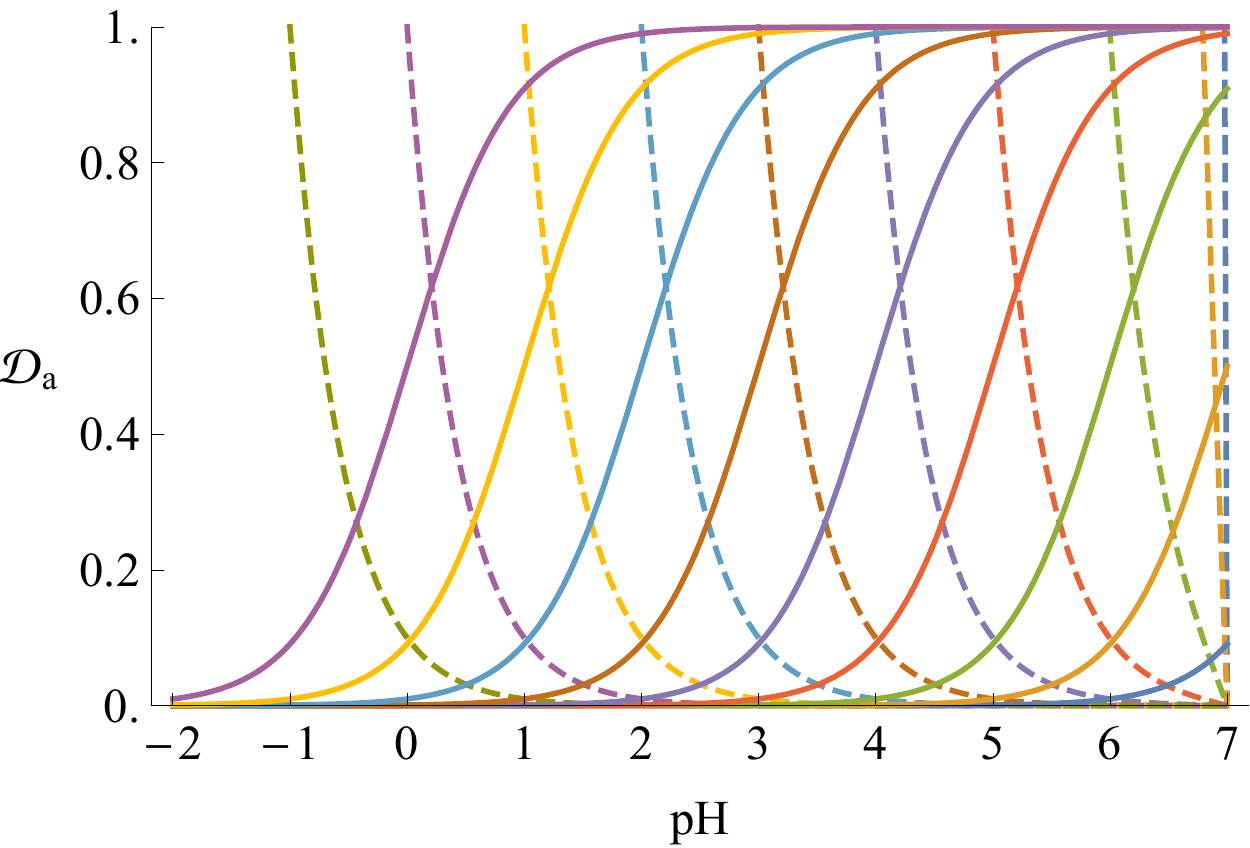}
    \caption{Constant $C_\mathrm{a}$ (dashed) and $K_\mathrm{a}$ (solid) curves as functions of $\mathrm{pH}$ and $\mathcal{D}_\mathrm{a}$. The dashed curves go from left to right, from the highest concentration $C_\mathrm{a}=10\,\mathrm{M}$ on the left (green), to lowest concentration $C_\mathrm{a}=10^{-9}\,\mathrm{M}$, on the right (blue). The solid curves go from left to right, from the stronger acid $K_\mathrm{a}=1\,\mathrm{M}$ on the left (purple), to the weakest acid $K_\mathrm{a}=10^{-8}\,\mathrm{M}$, on the right (blue).}
    \label{fig:ca_ka_pH_da}
\end{figure}

The use of $\rho=10^{7-\mathrm{pH}}$ in equations \eqref{eq:degree_of_dissociation_1} and \eqref{eq:ka_in_terms_of_da_and_rho} give the functions $c_\mathrm{a}(\mathrm{pH},\mathcal{D}_\mathrm{a})$ and $k_\mathrm{a}(\mathrm{pH},\mathcal{D}_\mathrm{a})$. Figure \ref{fig:ca_ka_pH_da} displays curves of constant $C_\mathrm{a}$ and $K_\mathrm{a}$ as functions of the pH and the degree of dissociation, $\mathcal{D}_\mathrm{a}$. The curves for $K_\mathrm{a}=\sqrt{K_\mathrm{w}}k_\mathrm{a}$ constant have sigmoid shape, increasing monotonically with the pH and reaching the highest value of $\mathcal{D}_\mathrm{a}$ at $\mathrm{pH}=7$. The stronger acids reach almost complete dissociation at $\mathrm{pH}=7$, meanwhile the weakest acids reach a dissociation of $\mathcal{D}_\mathrm{a}\approx 0.1$ at $\mathrm{pH}=7$.

Equation \eqref{eq:ka_in_terms_of_da_and_rho} can be rearranged to have $\mathcal{D}_\mathrm{a}(\mathrm{pH},k_\mathrm{a})$,
\begin{equation}
    \mathcal{D}_\mathrm{a}=\left(1+\frac{10^{7-\mathrm{pH}}}{k_\mathrm{a}}\right)^{-1},
\end{equation}
The sigmoid function $\mathcal{D}_\mathrm{a}(\mathrm{pH},k_\mathrm{a})$ is bounded and differentiable at all values of $\mathrm{pH}$ and $k_\mathrm{a}$, and monotonically non-decreasing with respect to $\mathrm{pH}$. The derivative of $\mathcal{D}_\mathrm{a}(\mathrm{pH},k_\mathrm{a})$ with respect to pH gives
\begin{align}
    \mathcal{P}_\mathrm{a}(\mathrm{pH},k_\mathrm{a})&=\left(\frac{\partial \mathcal{D}_\mathrm{a}}{\partial \mathrm{pH}}\right)\\
    &=\frac{\ln{10}}{k_\mathrm{a}}\frac{10^{7-\mathrm{pH}}}{\left(1+\frac{10^{7-\mathrm{pH}}}{k_\mathrm{a}}\right)^{2}}.
\end{align}
The maximum of $\mathcal{P}_\mathrm{a}$ is given by the solution of $\left({\partial\mathcal{P}_\mathrm{a}}/{
\partial\mathrm{pH}}\right)=0$, which is given by ${\mathrm{pH}^*=7-\log_{10}{k_\mathrm{a}}}$,  $\mathrm{pH}^*=-\log_{10}{K_\mathrm{a}}$, or $\mathrm{pH}^*=\mathrm{p}K_\mathrm{a}$.

\subsection{Titration by a strong base}

\begin{figure}[htb]
    \centering
    \includegraphics[scale=0.7]{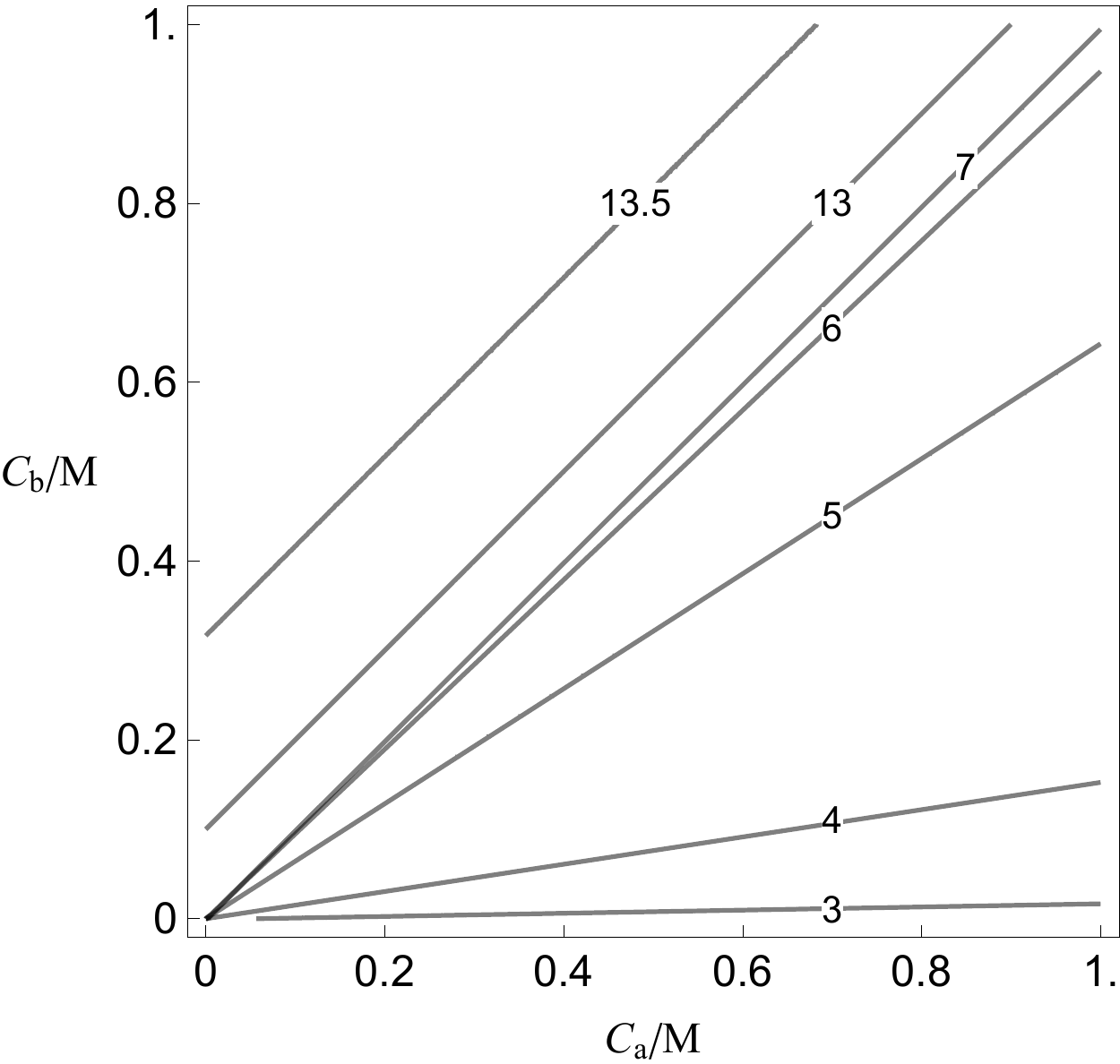}
    \caption{Lines of constant $\mathrm{pH}$ on the $C_\mathrm{a}$-$C_\mathrm{b}$ plane.}
    \label{fig:pH_acid_base}
\end{figure}

The addition of a strong base, \emph{e.g.} $\ce{NaOH}$, has direct effect on the charge balance \eqref{eq:charge_balance} due to the complete dissociation $\ce{NaOH -> Na+ + OH-}$. In terms of dimensionless concentrations, the equilibrium equations for the mixture of a weak acid and a strong base are \cite{Denbigh1981,Burgot2012,Skoog2022}
\begin{align}
    k_\mathrm{a}&=\frac{xz}{w},\label{eq:ka2}\\
    k_\mathrm{w}&=xy=1,\label{eq:kw2}\\
    x+s&=y+z,\label{eq:cb2}\\
    c_\mathrm{a}&=w+z,\label{eq:c2}\\
    c_\mathrm{b}&=s.\label{eq:cbase2}
\end{align}
In this equations $s=\ce{[Na+]}/\sqrt{K_\mathrm{w}}$, and $c_\mathrm{b}=C_\mathrm{b}/\sqrt{K_\mathrm{w}}$, are the concentrations of sodium ions and the base respectively. Algebraic manipulation of equations \eqref{eq:ka2}--\eqref{eq:cbase2} give the polynomial equation $P_\mathrm{at}(x)=0$, with
\begin{equation}
    P_\mathrm{at}(x)=x^3+\left(k_\mathrm{a}+c_\mathrm{b}\right)x^2-\left(1+k_\mathrm{a}\left(c_\mathrm{a}-c_\mathrm{b}\right)\right)x-k_\mathrm{a}.
\end{equation}
The use of Wolfram Mathematica \cite{Mathematica} allows to prove that the discriminant of $P_\mathrm{at}(x)$ is greater than zero, $\Delta[P_\mathrm{at}]>0$, hence there are three real solutions. It can be proved that only one of these roots is positive by following the same algebraic analysis that led to equation \eqref{eq:rho3_q_negative} from $P_\mathrm{a}(x)=0$. The positive root is given by
\begin{equation}\label{eq:roots_acid_base}
    \rho=2\sqrt[3]{\|\zeta\|}\cos{\left(\frac{\theta}{3}\right)-\frac{k_\mathrm{a}+c_\mathrm{b}}{3}},
\end{equation}
with
\begin{align}
    \sqrt[3]{\|\zeta\|}&=\tfrac{1}{3}\sqrt{\left(k_\mathrm{a}+c_\mathrm{b}\right)^2+3k_\mathrm{a}\left(c_\mathrm{a}-c_\mathrm{b}\right)+3},\label{eq:zeta_2}\\
    \theta&=\arctan{\left(-\frac{q}{2},\frac{\sqrt{\Delta[P_\mathrm{at}]}}{6\sqrt{3}}\right)},\label{eq:theta_def2}\\
    \Delta[P_\mathrm{at}]&=-4p^3-27q^2,\label{eq:disc_def2}\\
    p&=-\tfrac{1}{3}\left(k_\mathrm{a}+c_\mathrm{b}\right)^2-k_\mathrm{a}\left(c_\mathrm{a}-c_\mathrm{b}\right)-1,\label{eq:p_def2}\\
    q&=\frac{2}{27}\left(k_\mathrm{a}+c_\mathrm{b}\right)^3+\frac{k_\mathrm{a}}{3}\left(k_\mathrm{a}+c_\mathrm{b}\right)\left(c_\mathrm{a}-c_\mathrm{b}\right)-\frac{2k_\mathrm{a}}{3}+\frac{c_\mathrm{b}}{3}.\label{eq:q_def_2}
\end{align}
The contours of $\mathrm{pH}=7-\log_{10}{\rho}$ as function of $C_\mathrm{a}=\sqrt{K_\mathrm{w}}c_\mathrm{a}$ and $C_\mathrm{b}=\sqrt{K_\mathrm{w}}c_\mathrm{b}$ for acetic acid, $k_\mathrm{a}=180$, are shown in figure \ref{fig:pH_acid_base}. It is interesting noticing that there is a linear relationship between base and acid concentrations at constant $\mathrm{pH}$. The equivalence point is given by the ${\mathrm{pH}=7}$ line, it has slope one and passes through the origin ${(c_\mathrm{a},c_\mathrm{b})=(0,0)}$. Figure \ref{fig:pH_acid_base} also shows that the $C_\mathrm{b}$-intercept is negative for acidic $\mathrm{pH}$ and positive for basic $\mathrm{pH}$. It is also seen in this figure, that the slope of the lines $C_\mathrm{b}(C_\mathrm{a})$ is less than one for acidic $\mathrm{pH}$.

The titration experiment consists of the addition of a volume $V_\mathrm{b}$ of a strong base solution, with analytical concentration $c_\mathrm{b}^{0}$, to a volume $V_\mathrm{a}$ of a weak acid solution, with analytical concentration $c_\mathrm{a}^0$. As the strong base is added the pH of the mixture increases. The equivalence point is reached when the moles of the acid and the base are equal, \emph{i.e.}, when the concentration of the acid and the base are equal in the mixture. The relation between the mixture concentrations and the analytical concentrations are given by
\begin{align}
    c_\mathrm{a}&=\frac{c_\mathrm{a}^0V_\mathrm{a}}{V_\mathrm{a}+V_\mathrm{b}},\\
    c_\mathrm{b}&=\frac{c_\mathrm{b}^0V_\mathrm{b}}{V_\mathrm{a}+V_\mathrm{b}}.
\end{align}
These equations can be combined to obtain
\begin{equation}
    c_\mathrm{b}=c_\mathrm{b}^0\left(1-\frac{c_\mathrm{a}}{c_\mathrm{a}^0}\right),
\end{equation}
which is the equation of a straight line with slope $-c_\mathrm{b}^0/c_\mathrm{a}^0$ and intercept $c_\mathrm{b}^0$.

\begin{figure}[htb]
    \centering
    \includegraphics[scale=0.7]{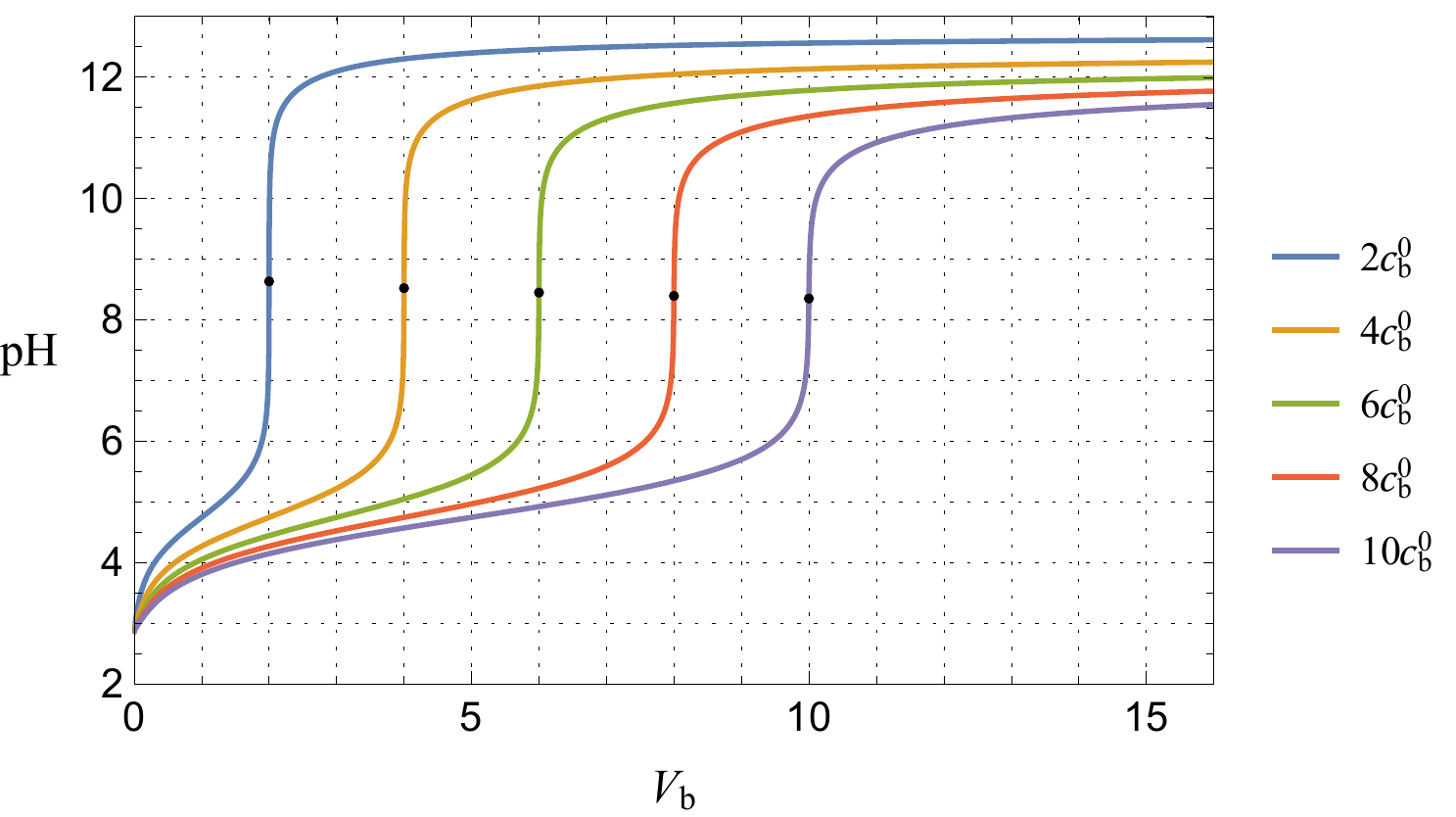}
    \caption{Titration curves for $V_\mathrm{a}=1\,\mathrm{L}$ of acetic acid solution, $k_\mathrm{a}=180$, with concentration $c_\mathrm{a}^0=0.1/\sqrt{K_\mathrm{w}}$. The concentration of the base is taken such that $n c_\mathrm{b}^0=c_\mathrm{a}^0$, with $n=2,4,6,8, 10$. The black point on each titration curve indicates the condition $c_\mathrm{a}=c_\mathrm{b}$. }
    \label{fig:titration_acetic_acid}
\end{figure}

The volume of the acid $V_\mathrm{a}$ is fixed from the beginning of the titration experiment. Figure \ref{fig:titration_acetic_acid} displays titration curves for $V_\mathrm{a}=1\,\mathrm{L}$ of an acetic acid solution, $k_\mathrm{a}=180$, with concentration $c_\mathrm{a}^0=0.1/\sqrt{K_\mathrm{w}}$. The concentration of the base has been chosen such that $nc_\mathrm{b}^0=c_\mathrm{a}^0$ with $n=2,4,6,8,10$. All the titration curves of figure \ref{fig:titration_acetic_acid} are monotonically increasing functions of the base volume $V_\mathrm{b}$, with the same $\mathrm{pH}$ at $V_\mathrm{b}=0$, $\mathrm{pH}=2.37$. It is clear that the more concentrated the base, the faster the $\mathrm{pH}$ grows as function of $V_\mathrm{b}$. 

The acid-base equivalence point is given when the titration curves reach the neutral $\mathrm{pH}$, $\mathrm{pH}=7$. The equivalence point of the titration curves of figure \ref{fig:titration_acetic_acid} are displayed in table \ref{table:table_1}. The analytical condition $c_\mathrm{a}=c_\mathrm{b}$, is reached at the black points indicated on each titration curve on figure \ref{fig:titration_acetic_acid}. The volume of added base $V_\mathrm{b}$, and the $\mathrm{pH}$ at the condition $c_\mathrm{a}=c_\mathrm{b}$ is displayed in table \ref{table:table_1}.

\begin{table}[htb]
\begin{center}
\begin{tabular}{|c | c | c | c|} 
\hline
 $n$ ($nc_\mathrm{b}^0=c_\mathrm{a}^0$) & $V_\mathrm{b}$ ($\mathrm{pH}=7$) & $V_\mathrm{b}$ ($c_\mathrm{a}=c_\mathrm{b}$) & $\mathrm{pH}$ ($c_\mathrm{a}=c_\mathrm{b}$)  \\  
 \hline
 2 & 1.989 & 2 & 8.634 \\ 
 4 & 3.978 & 4 & 8.523\\
 6 & 5.967 & 6 & 8.450\\
 8 & 7.956 & 8 & 8.396 \\
 10 & 9.945 & 10 & 8.352 \\ 
 \hline
\end{tabular}
\caption{Volume of the base, $V_\mathrm{b}$, at the equivalence point ($\mathrm{pH}=7$), at the equal concentrations condition ($c_\mathrm{a}=c_\mathrm{b}$), and $\mathrm{pH}$ at the concentration condition ($c_\mathrm{a}=c_\mathrm{b}$).}
\label{table:table_1}
\end{center}
\end{table}

\subsection{Buffer solution}

An acid buffer solution is the aqueous dissolution of a weak acid ($\ce{HB}$) and a fully soluble salt of the conjugate base ($\ce{NaB}$). The chemical equations of a buffer equilibrium are \cite{Denbigh1981,Burgot2012,Skoog2022}
\begin{align}
    \ce{HB + H2O &<--> H3O+ + B-},\label{ce:HA_diss_buffer}\\
    \ce{B- + H2O &<--> HB + OH-},\label{ce:B_hydrolysis_buffer}\\
    \ce{NaB &-> Na+ + B-}\\
    \ce{2 H2O &<--> H3O+ + OH-}.\label{ce:H2O_auto_buffer}
\end{align}
Relevant chemical species are $\ce{H3O+}$, $\ce{OH-}$, $\ce{HB}$, $\ce{B-}$, and $\ce{Na+}$, with molar concentrations  $\ce{[H3O+]}$, $\ce{[OH^-]}$, $\ce{[HB]}$, $\ce{[B^-]}$, and $\ce{[Na+]}$. 

In an acid buffer solution with acid concentration $C_\mathrm{a}$ and salt concentration $C_\mathrm{s}$, the concentrations  of the chemical species are given by four physical conditions: the weak acid dissociation constant $K_\mathrm{a}$, the water auto-ionization constant $K_\mathrm{w}$, the electric neutrality, and the mass balance,
\begin{align}
    K_\mathrm{a}&=\frac{\ce{[H3O+]}\ce{[B^-]}}{\ce{[HB]}},\label{eq:Ka_buffer}\\
    K_\mathrm{w}&=\ce{[H3O^+]}\ce{[OH^-]},\label{eq:Kw_buffer}\\
    \ce{[H3O+]} + \ce{[Na+]} &=\ce{[OH^-]}+\ce{[B^-]},\label{eq:charge_balance_buffer}\\
    C_\mathrm{a} +  C_\mathrm{s}&=\ce{[B^-]}+\ce{[HB]},\label{eq:matter_balance_buffer}
\end{align}
respectively. Since the salt $\ce{NaB}$ dissociates completely, the concentration of $\ce{Na+}$ is given by $C_\mathrm{s}=\ce{[Na^+]}$. The hydrolysis of the conjugate base, chemical equation \eqref{ce:B_hydrolysis_buffer}, gives the conjugate base hydrolysis constant $K_\mathrm{cb}=K_{\mathrm{w}}/K_{\mathrm{a}}$, which is irrelevant for this analysis.

Acid constants $K_\mathrm{a}$ have units of concentration and their value range typically between $10^{-10}$ and $10^{-1}$ $\mathrm{M}$. It is convenient to define the dimensionless variables and constants ${s=\ce{[Na^+]}/\sqrt{K_\mathrm{w}}}$, ${x=\ce{[H3O^+]}/\sqrt{K_\mathrm{w}}}$, ${y=\ce{[OH^-]}/\sqrt{K_\mathrm{w}}}$, ${w=\ce{[HB]}/\sqrt{K_\mathrm{w}}}$, ${z=\ce{[B^-]}/\sqrt{K_\mathrm{w}}}$, ${c_\mathrm{a}=C_\mathrm{a}/\sqrt{K_\mathrm{w}}}$, ${c_\mathrm{s}=C_\mathrm{s}/\sqrt{K_\mathrm{w}}}$, ${k_a=K_\mathrm{a}/\sqrt{K_\mathrm{w}}}$, and ${k_\mathrm{w}=1}$. In terms of the dimensionless variables and constants, equations \eqref{eq:Ka_buffer}-\eqref{eq:matter_balance_buffer} are replaced by
\begin{align}
    k_\mathrm{a}&=\frac{xz}{w},\label{eq:ka_buffer}\\
    k_\mathrm{w}&=xy=1,\label{eq:kw_buffer}\\
    x+c_\mathrm{s}&=y+z,\label{eq:cb_buffer}\\
    c_\mathrm{a}+c_\mathrm{s}&=w+z.\label{eq:c_buffer}
\end{align}

The addition of a strong base, \emph{e.g.} $\ce{NaOH -> Na+ + OH-}$, affects the charge balance, equation \eqref{eq:cb_buffer}. This addition is useful to study the buffer capacity of the solution. The use of a strong base with analytical concentration ${C_\mathrm{b}=\sqrt{K_\mathrm{w}}c_\mathrm{b}}$, gives the charge balance 
\begin{equation}
        x+c_\mathrm{s}+c_\mathrm{b}=y+z.\label{eq:cb_buffer_2}\\
\end{equation}

The combined use of equations \eqref{eq:ka_buffer}-\eqref{eq:cb_buffer_2} gives
\begin{equation}\label{eq:ka_x_buffer}
    k_\mathrm{a}=\frac{x\left(x+c_\mathrm{s}+c_\mathrm{b}-\tfrac{1}{x}\right)}{\tfrac{1}{x}-x+c_\mathrm{a}-c_\mathrm{b}}.
\end{equation}
Since $x>0$, and $k_\mathrm{a}>0$, the numerator and denominator of \eqref{eq:ka_x_buffer} give two quadratic constraints for the concentration $x$,
\begin{align}
    x^2+\left(c_\mathrm{s}+c_\mathrm{b}\right)x-1&>0,\label{eq:quadratic_constraint_buffer_1}\\
    -x^2+\left(c_\mathrm{a}-c_\mathrm{b}\right)x+1&>0,\label{eq:quadratic_constraint_buffer_2}
\end{align}
respectively.


Rearrangement of equation \eqref{eq:ka_x_buffer} leads to the polynomial equation $P_\mathrm{ab}(x)=0$, with 
\begin{equation}\label{eq:P_polynomial_buffer}
    P_\mathrm{ab}(x)=x^3+\left(k_\mathrm{a}+c_\mathrm{b}+c_\mathrm{s}\right)x^2-\left(1+k_\mathrm{a}\left(c_\mathrm{a}-c_\mathrm{b}\right)\right)x-k_\mathrm{a}.
\end{equation}
The dimensionless equation $P_\mathrm{ab}(x)=0$ is equivalent to Charlot's equation when ${c_\mathrm{b}=0}$.

The use of Wolfram Mathematica allows to prove that the discriminant of $P_\mathrm{at}(x)$ is greater than zero, $\Delta[P_\mathrm{ab}]>0$, hence there are three real solutions. It can be proved that only one of these roots is positive by following the same algebraic analysis that led to equation \eqref{eq:rho3_q_negative} from $P_\mathrm{a}(x)=0$. The positive root, ${\rho=\rho(c_\mathrm{a},c_\mathrm{b},c_\mathrm{s})}$, is given by
\begin{equation}\label{eq:roots_acid_buffer}
        \rho=2\sqrt[3]{\|\zeta\|}\cos{\left(\frac{\theta}{3}\right)-\frac{k_\mathrm{a}+c_\mathrm{b}+c_\mathrm{s}}{3}}, 
\end{equation}
with
\begin{align}
    \sqrt[3]{\|\zeta\|}&=\tfrac{1}{3}\sqrt{\left(k_\mathrm{a}+c_\mathrm{b}+c_\mathrm{s}\right)^2+3k_\mathrm{a}(c_\mathrm{a}-c_\mathrm{b})+3},\label{eq:zeta_3}\\
    \theta&=\arctan{\left(-\frac{q}{2},\frac{\sqrt{\Delta[P_\mathrm{ab}]}}{6\sqrt{3}}\right)},\label{eq:theta_def3}\\
    \Delta[P_\mathrm{ab}]&=-4p^3-27q^2,\label{eq:disc_def3}\\
    p&=-\tfrac{1}{3}\left(k_\mathrm{a}+c_\mathrm{b}+c_\mathrm{s}\right)^2-k_\mathrm{a}\left(c_\mathrm{a}-c_\mathrm{b}\right)-1,\label{eq:p_def3}\\
     q&=\tfrac{2}{27}\left(k_\mathrm{a}+c_\mathrm{b}+c_\mathrm{s}\right)^3+\frac{k_\mathrm{a}}{3}\left(k_\mathrm{a}+c_\mathrm{b}+c_\mathrm{s}\right)\left(c_\mathrm{a}-c_\mathrm{b}\right)-\frac{2k_\mathrm{a}}{3}+\frac{c_\mathrm{b}+c_\mathrm{s}}{3}.\label{eq:q_def_3}
\end{align}

\begin{figure}
    \centering
    \includegraphics[scale=0.333]{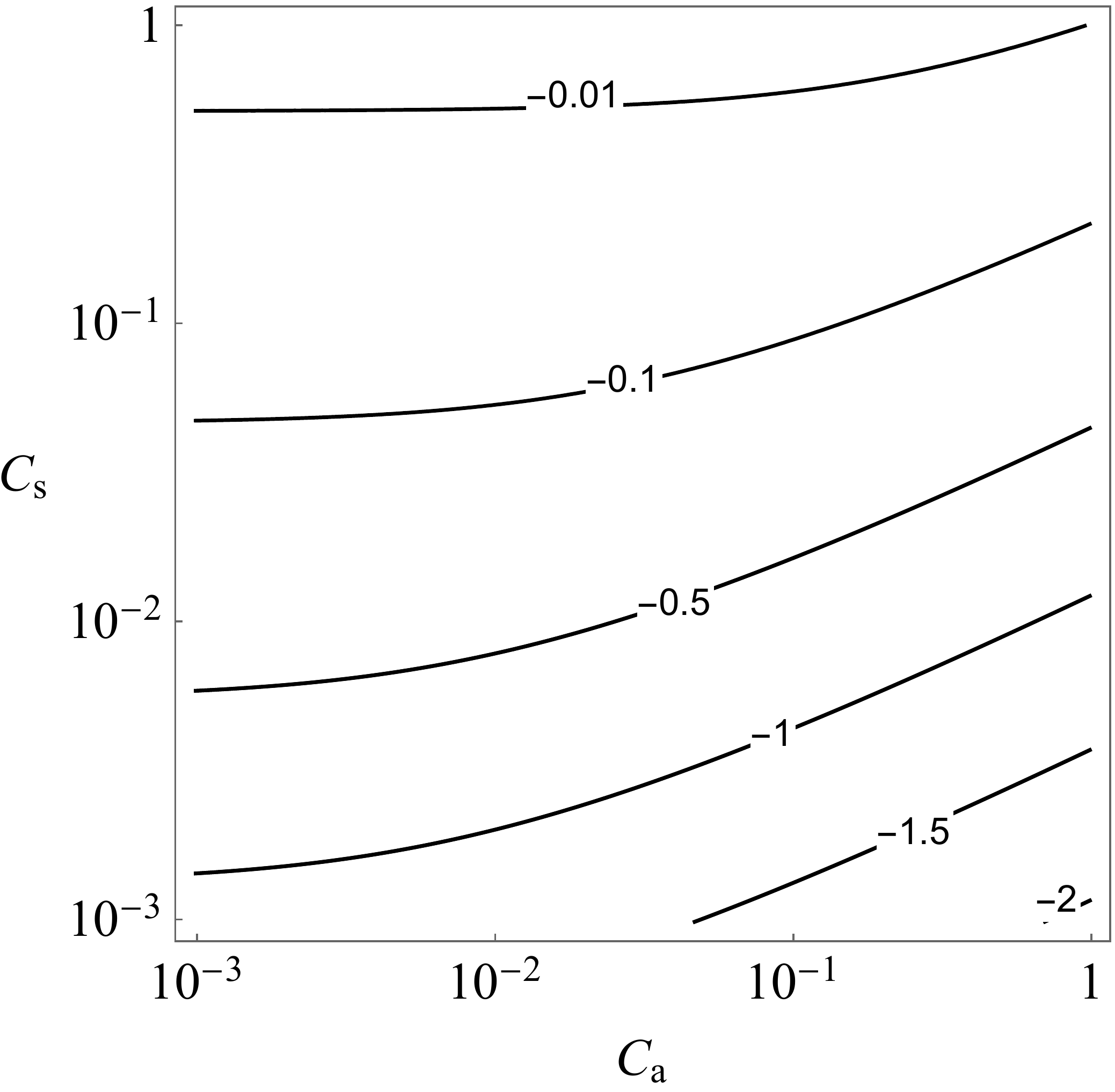}
    \caption{Abslute error $E^{\mathrm{(pH)}}$, in the $\mathrm{pH}$ calculated by the Herderson-Hasselbach equation for a buffer solution of chlorous acid, $K_\mathrm{a}=1.2\times 10^{-2}\,\mathrm{M}$, as a function of the molar concentrations of the acid $C_\mathrm{a}$ and the salt $C_\mathrm{s}$.}
    \label{fig:HH_approx}
\end{figure}

By applying the logarithm of the acid dissociation constant $K_\mathrm{a}$, equation \eqref{eq:Ka_buffer}, is obtained, after some algebraic manipulation, the Henderson-Hasselbach (HH) equation \cite{Henderson1908,Hasselbalch1917,Po2001}
\begin{equation}\label{eq:HH}
    \mathrm{pH}=\mathrm{p}K_\mathrm{a}+\log_{10}{\frac{\ce{[B^-]}}{\ce{[HB]}}}.
\end{equation}
The right hand side of this equation is a function of $\mathrm{pH}$, therefore the HH equation is not practical for direct calculation of the $\mathrm{pH}$. It is common to use the approximations $\ce{[B^-]}\approx C_\mathrm{s}$ and $\ce{[HB]}\approx C_\mathrm{a}$ to obtain
\begin{equation}\label{eq:HH_approx}
    \mathrm{pH}^{\mathrm{(1)}}\approx\mathrm{p}K_\mathrm{a}+\log_{10}{\frac{C_\mathrm{s}}{C_\mathrm{a}}}.
\end{equation}
Figure \ref{fig:HH_approx} shows the absolute error of the $\mathrm{pH}$, for chlorous acid, calculated by the HH equation with respect to the pH calculated using the exact formula, \eqref{eq:roots_acid_buffer} with $c_\mathrm{b}=0$,
\begin{equation}
    E^{\mathrm{(pH)}}=\mathrm{pH}^{\mathrm{(1)}}-\left(7-\log_{10}{\rho}\right),
\end{equation}
with $\rho$ given by equation \eqref{eq:roots_acid_buffer}. Figure \ref{fig:HH_approx} shows that small errors, $E^\mathrm{(pH)}<0.1$, are obtained for buffer solutions with high concentration of the salt, $C_\mathrm{s}>0.1\,\mathrm{M}$, meanwhile large errors, $E^\mathrm{(pH)}>0.5$ are obtained for buffer solutions with $C_\mathrm{s}<0.01\,\mathrm{M}$. 

\subsubsection{Buffer capacity}

The pH stability of an acid buffer solution is measured by adding a volume $V_\mathrm{b}$ of a strong base solution with concentration $c_\mathrm{b}^0$ \cite{Skoog2022}. The addition of this volume changes the concentrations $c_\mathrm{a}$, $c_\mathrm{s}$, and $c_\mathrm{b}$,
\begin{align}
    c_{\mathrm{a}}&=\frac{c_{\mathrm{a}}^0 V_{\mathrm{a}}^0}{V_{\mathrm{ab}}^0+V_{\mathrm{b}}},\label{eq:ca_Vb}\\
    c_{\mathrm{s}}&=\frac{c_\mathrm{s}^0 V_{\mathrm{s}}^0}{V_{\mathrm{ab}}^0+V_{\mathrm{b}}},\label{eq:cs_Vb}\\ 
    c_{\mathrm{b}}&=\frac{c_\mathrm{b}^0 V_{\mathrm{b}}}{V_{\mathrm{ab}}^0+V_{\mathrm{b}}},\label{eq:cb_Vb}
\end{align}
with $c_\mathrm{a}^0$, $c_\mathrm{s}^0$, and $c_\mathrm{b}^0$, as the concentrations of the acid, salt of acid, and base independent solutions, respectively. The volumes $V_\mathrm{a}^0$, $V_\mathrm{s}^0$, and $V_\mathrm{ab}^0=V_\mathrm{a}^0+V_\mathrm{s}^0$, are the initial volumes of the  acid, salt, and acid buffer solutions respectively.

The pH stability, $S_\mathrm{pH}$, of a buffer solution is given by \begin{equation}\label{eq:pH_stability}
\begin{split}
    S_{\mathrm{pH}}&=\frac{d\left(\mathrm{pH}\right)}{d V_\mathrm{b}}\\
    &=\nabla_{\bm{c}}(\mathrm{pH})\cdot\frac{d\bm{c}}{d V_b},
\end{split}
\end{equation}
with $\bm{c}=\left({c_\mathrm{a}},{c_\mathrm{s}},{c_\mathrm{b}}\right)$ as 3-vector of concentrations, and the gradient of $\rho$, $\nabla_{\bm{c}}\rho$, given by
\begin{equation}
    \nabla_{\bm{c}}\rho=\left(\frac{\partial\rho}{\partial c_\mathrm{a}},\frac{\partial\rho}{\partial c_\mathrm{s}},\frac{\partial\rho}{\partial c_\mathrm{b}}\right).
\end{equation}
The use in equation \eqref{eq:pH_stability} of the pH definition, $\mathrm{pH}=7-{\log_{10}}\rho$, and the concentrations of equations \eqref{eq:ca_Vb}-\eqref{eq:cb_Vb}, gives after some algebra 
\begin{equation}
    S_\mathrm{pH}=\frac{1}{{\ln{10}}}\frac{1}{\left(V_\mathrm{ab}^0+V_\mathrm{b}\right)^2}\frac{1}{\rho}{\nabla_{\bm{c}}}\rho\cdot{\bm{n}^0},
\end{equation}
with
\begin{equation}
\bm{n}^0=\begin{pmatrix} c_\mathrm{a}^0V_\mathrm{a}^0, & c_\mathrm{s}^0V_\mathrm{s}^0, & -c_\mathrm{b}^0V_\mathrm{ab}^0\end{pmatrix}.
\end{equation}

The gradient $\nabla_{\bm{c}}\rho$ is calculated with respect to the components of $\bm{c}$ but it must be expressed in terms of the volume of base $V_\mathrm{b}$.

\begin{figure}
    \centering
    \includegraphics[scale=0.5]{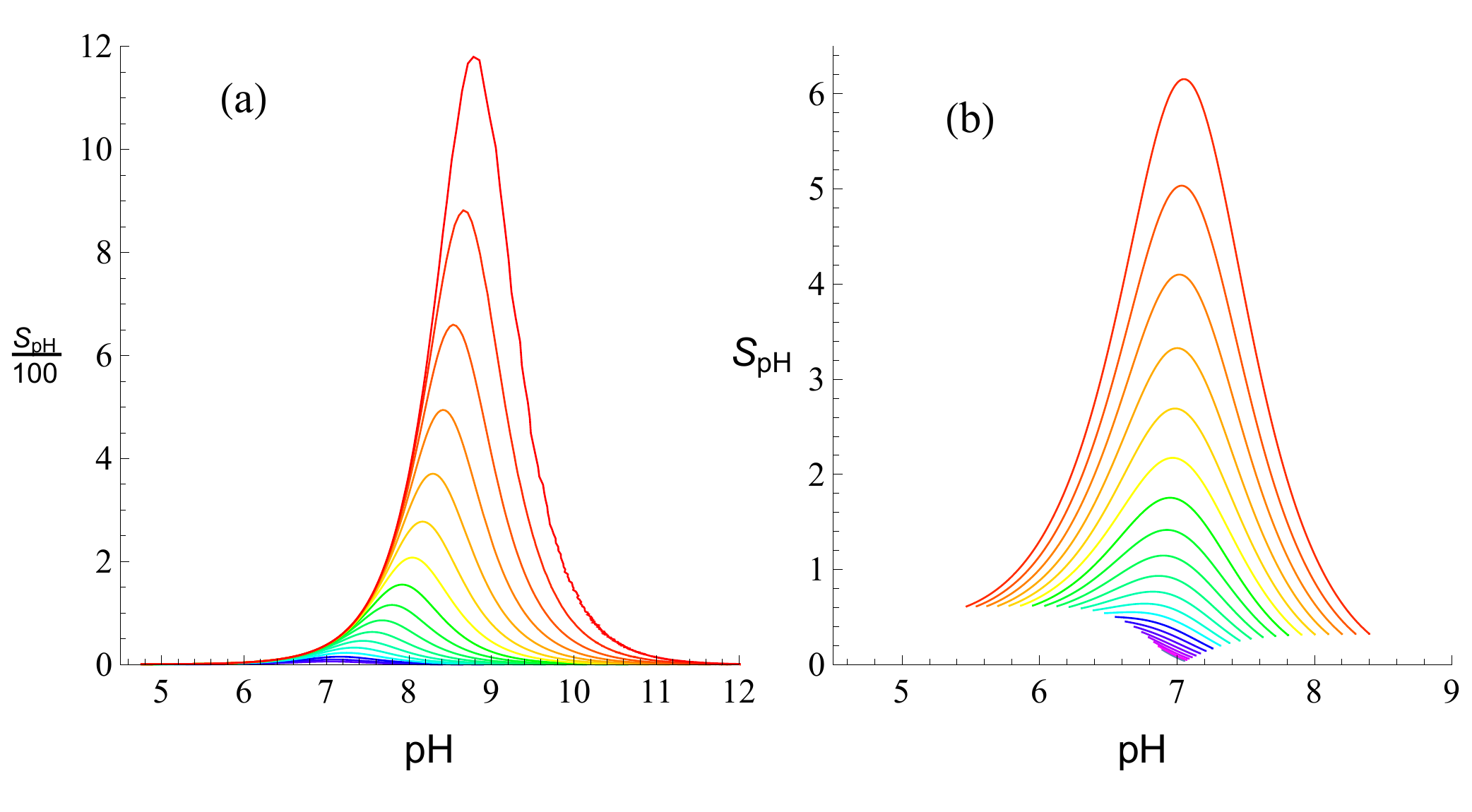}
    \caption{$\mathrm{pH}$ stability $S_\mathrm{pH}$ as a function of the pH for buffer solutions of acetic acid ($k_\mathrm{a}=175$) and sodium acetate titrated by NaOH. To simplify the analysis, the initial concentrations $c_\mathrm{a}^0$, $c_\mathrm{s}^0$, and $c_\mathrm{b}^0$, are equal to $c^0$. Panel (a) displays $S_\mathrm{pH}/100$ for solutions with concentrations $C^0=10^{-7}\times c^0$ from $10^{-5}$ to $10^{-1}\,\mathrm{M}$; The lowest curves (blue color) are at lower concentrations, the highest curves (red color) are for higher concentrations. Panel (b) displays $S_\mathrm{pH}$ for solutions with concentrations $C^0=10^{-7}\times c^0$ from $10^{-7}$ to $10^{-5}\,\mathrm{M}$; The lowest curves (purple color) are for low concentration buffers, the highest curves (red color) for higher concentrations. }
    \label{fig:buffer_stability}
\end{figure}

Figure \ref{fig:buffer_stability} displays the $\mathrm{pH}$ stability, $S_\mathrm{pH}$, as function of the $\mathrm{pH}$ of the buffer solution. Since the addition of a strong base increases monotonically the $\mathrm{pH}$ of the solution, the curves of figure \ref{fig:buffer_stability} contain the same information as the titration curves. To simplify the analysis all the concentrations used to prepare or titrate the buffers, $c_\mathrm{a}^0$, $c_\mathrm{s}^0$ and $c_\mathrm{b}^0$, have the same value $c^0$. Panel (a) of figure \ref{fig:buffer_stability} displays $S_\mathrm{pH}/100$ for buffer solutions with concentrations $C^0=10^{-7}\times c^0$ ranging from $10^{-2}$ to $10^{-1}\,\mathrm{M}$ (from blue to red). Panel (b) of figure \ref{fig:buffer_stability} displays $S_\mathrm{pH}$ for buffer solutions with concentrations $C^0=10^{-7}\times c^0$ ranging from $10^{-7}$ to $10^{-5}\,\mathrm{M}$ (from purple to re). In both panels of figure \ref{fig:buffer_stability}, curves with lower (higher) values of $S_\mathrm{pH}$ are for buffer solutions prepared and titrated with solutions of lower (higher) concentrations. The maximum of the curves $S_\mathrm{pH}$ is higher, and located at higher pH, for buffer solutions at higher concentrations $C^0$. The highest concentration buffer of figure \ref{fig:buffer_stability}(a), in red) displays numerical instability for basic $\mathrm{pH}s$. 

\begin{figure}
    \centering
    \includegraphics[scale=0.5]{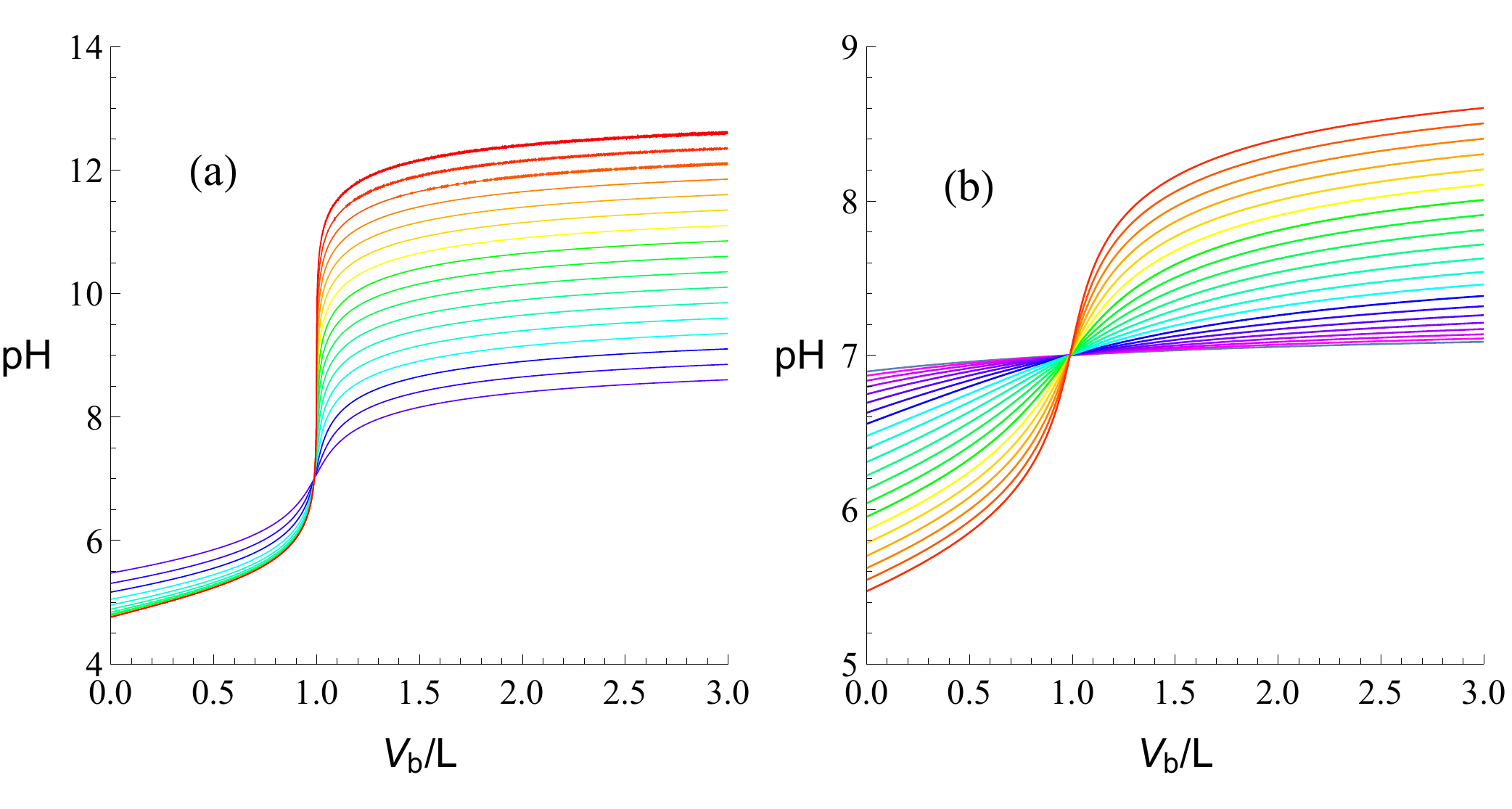}
    \caption{Buffer titration as function of the volume of base added, $V_\mathrm{b}/\mathrm{L}$. To simplify the analysis, the concentrations of the solutions to prepare, and titrate, the buffer are the same, $C^0=10^{-7}c^0$. Panel (a) displays the $\mathrm{pH}$ for solutions with concentrations $C^0=10^{-7}\times c^0$ from $10^{-5}$ to $10^{-1}\,\mathrm{M}$; The blue blue curves are at lower concentrations, the red curves are for higher concentrations. Panel (b) displays $\mathrm{pH}$ for solutions with concentrations $C^0=10^{-7}\times c^0$ from $10^{-7}$ to $10^{-5}\,\mathrm{M}$; The purple curves are for low concentration buffers, the red curves for higher concentrations. }
    \label{fig:buffer_titration}
\end{figure}

Figure \ref{fig:buffer_titration} displays the titration curves for the same buffer solutions of figure \ref{fig:buffer_stability}. To simplify the analysis, the concentrations of the solutions to prepare, and titrate, the buffer are the same, $C^0=10^{-7}c^0$. All the titration curves of panels (a) and (b) of figure \ref{fig:buffer_titration} have the same equilibrium point at $V_\mathrm{b}=1\,\mathrm{L}$ and $\mathrm{pH}=7$. Buffers of higher concentrations, in red, display the largest changes of $\mathrm{pH}$ as the solution is titrated with strong base. 

The use of figures \ref{fig:buffer_stability}(a) and \ref{fig:buffer_titration}(a) indicates that the lower the concentration of the solutions of acid and salt of the acid (blue curves), the lower the change in $\mathrm{pH}$ as the base is added. 

\section{Systematic analysis relating the pH expressions}

The four cubic equations for $x={\ce{[H3O+]}}/{\sqrt{K_\mathrm{w}}}$ of the systems under study can be written as
\begin{equation}\label{eq:general_polynomial}
    x^3+a_2 x^2 +a_1 x+a_0=0.
\end{equation}
The coefficients of the four systems under study are given by the rows of table \ref{tab:coeff_syst}

\begin{table}[hb]
    \centering
    \begin{tabular}{|l|c|c|c|c|}
    \hline
       System & $a_2=k$ & $a_1$ & $a_0$ & $c$ \\
       \hline
       w.a.  & $k_\mathrm{a}$ & $-c_\mathrm{a}k_\mathrm{a}-1$ & $-k_\mathrm{a}$ & $c_\mathrm{a}$ \\
       w.a. + s.b.  & $k_\mathrm{a}+c_\mathrm{b}$ &  $-c_\mathrm{a}k_\mathrm{a}-1+c_\mathrm{b}k_\mathrm{a}$ & $-k_\mathrm{a}$ & $c_\mathrm{a}-c_\mathrm{b}$ \\
       w.a.b. & $k_\mathrm{a}+c_\mathrm{s}$  &  $-c_\mathrm{a}k_\mathrm{a}-1$ & $-k_\mathrm{a}$ & $c_\mathrm{a}$ \\
       w.a.b. + s.b.  & $k_\mathrm{a}+c_\mathrm{b}+c_\mathrm{s}$  &  $-c_\mathrm{a}k_\mathrm{a}-1+c_\mathrm{b}k_\mathrm{a}$ & $-k_\mathrm{a}$ & $c_\mathrm{a}-c_\mathrm{b}$\\
       \hline
    \end{tabular}
    \caption{Coefficients of the polynomials for the systems studied: weak acid (w.a.), weak acid titration by a strong base (w.a. + s.b.), weak acid buffer (w.a.b.) and titration of a weak acid buffer by a strong base (w.a.b. + s.b.). The second and fifth columns give the effective acid constant $k$ and concentration $c$, respectively.}
    \label{tab:coeff_syst}
\end{table}

The concentration $x$ is given by the only positive root of these polynomials,
\begin{equation}\label{eq:general_x}
    x=\tfrac{2}{3}\sqrt{k^2+3ck_a+3}\cos{\left(\theta/3\right)}-\tfrac{k}{3},
\end{equation}
with $c$ and $k$ as the effective concentrations and acid dissociation constants, given by the fifth and second columns of table \ref{tab:coeff_syst}, respectively. The angle $\theta$ is given by the expression
\begin{equation}\label{eq:general_theta}
    \theta=\arctan{\left(-\frac{q}{2},\frac{\sqrt{\Delta}}{6\sqrt{3}}\right)},
\end{equation}
with discriminant $\Delta=-4p^3-27q^2$, and
\begin{align}
    p&=-\tfrac{1}{3}k^2-k_\mathrm{a}c-1,\label{eq:general_p}\\
    q&=\tfrac{2}{27}k^3+\tfrac{k}{3}\left(1+k_\mathrm{a}c\right)-k_\mathrm{a}.\label{eq:general_q}
\end{align}
The values of $k$ and $c$ from table \ref{tab:coeff_syst} show that $k \ge k_\mathrm{a}$ and $c\le c_\mathrm{a}$. Equations \eqref{eq:general_x}--\eqref{eq:general_q} with the data of table \ref{tab:coeff_syst} give the full description of the pH for the systems studied.

\section{Conclusions}

A systematic algebraic analysis has been applied on the ideal aqueous chemical equilibrium of three acid-base systems of academic and practical interest. The weak acid dissociation, the titration of a weak acid solution by a strong base, the acid buffer solution and its titration by a strong base, have been algebraically analyzed without approximations. Simple analytical expression for the the roots of the cubic equations for the $\ce{[H3O+]}$ are obtained. It has been shown that all the systems under study have one real positive root and two negative roots, no complex roots were obtained. The existence of only one positive root, with physical meaning, was proved in several ways: by analyzing the discriminant of the cubic polynomials, by using Descates' rule of signs on the polynomial and its derivative, and by analyzing Vieta's formulas. The cubic equations for the systems of interest were solved by the method of the depressed cubic equation of Vieta and Cardano. It has been found that, for all the systems under study, $\ce{[H3O+]}$ is the subtraction of two terms; The first term is the product of a trigonometric function and the square-root of a quadratic form of the acid dissociation constant, $k_\mathrm{a}$, and the concentrations, $c_\mathrm{a}$, $c_\mathrm{s}$, and $c_\mathrm{b}$; The second term is simply one third of the sum of the acid constant and the concentrations. The  analytical forms of $\ce{[H3O+]}$ for the four systems analyzed are easily related by simple mathematical substitutions.

In addition to the expression for $\ce{[H3O+]}$, another analytical expression for the degree of dissociation of a weak acid $\mathcal{D}_\mathrm{a}$, the $pH$ titration curves for weak acids and buffer solutions, and the stability of the pH $S_\mathrm{pH}$ have been obtained and studied. Exact titration curves allow to predict accurately the neutralization of weak acids, and buffer solutions, by using strong bases. The expression obtained for the pH stability $S_\mathrm{pH}$ allows to quantify precisely the  pH stability of a buffer solution when a strong base is added. 

The exact $\mathrm{pH}$ has been compared against the approximate expression commonly used for calculating the $\mathrm{pH}$. The error of these approximations is analyzed in detail. It was found that the known rule of 5\% is based on assuming that ${\sqrt{K_\mathrm{w}}\ll\ce{[H3O+]}}$. It has been shown that the Henderson-Hasselbach (HH) equation exhibits large error for buffer solutions with salt concentrations  $C_\mathrm{s}>0.01\,\mathrm{M}$. 

\section*{Disclosure statement}

No potential conflict of interest was reported by the authors.

\section*{Funding}

This project has been fully financed by the internal research grants of the University.

\bibliography{references}

\end{document}